\documentclass[aps,pre,preprint,superscriptaddress]{revtex4-1}
\usepackage{graphicx,epsfig,color,textcomp,amssymb,amsmath,mathrsfs,cancel,bbm}

\usepackage{soul,color}
\soulregister\cite7
\soulregister\ref7
\soulregister\eqref7

\begin{document}
\title{Thermodynamic Relationships for Bulk Crystalline and Liquid Phases in the Phase-Field Crystal Model}
\date{\today}
\author{V. W. L. ~Chan}
\email{vicchan@umich.edu}
\affiliation{Materials Science and Engineering Department, University of Michigan, Ann Arbor, Michigan, 48109, USA}
\author{N. ~Pisutha-Arnond}
\email{kpnirand@kmitl.ac.th}
\affiliation{Faculty of Engineering, King Mongkut's Institute of Technology Ladkrabang, Bangkok, Thailand}
\author{K. ~Thornton}
\email{kthorn@umich.edu}
\affiliation{Materials Science and Engineering Department, University of Michigan, Ann Arbor, Michigan, 48109, USA}
\begin{abstract}
We present thermodynamic relationships between the free energy of the phase-field crystal (PFC) model and thermodynamic state variables for bulk phases under hydrostatic pressure. This relationship is derived based on the thermodynamic formalism for crystalline solids of Larch\'e and Cahn [Larch\'e and Cahn, Acta Metallurgica, Vol.\  21, 1051 (1973)]. We apply the relationship to examine the thermodynamic processes associated with varying the input parameters of the PFC model: temperature, lattice spacing, and the average value of the PFC order parameter, $\bar{n}$.  The equilibrium conditions between bulk crystalline solid and liquid phases are imposed on the thermodynamic relationships for the PFC model to obtain a procedure for determining solid-liquid phase coexistence. The resulting procedure is found to be in agreement with the method commonly used in the PFC community, justifying the use of the common-tangent construction to determine solid-liquid phase coexistence in the PFC model. Finally, we apply the procedure to an eighth-order-fit (EOF) PFC model that has been parameterized to body-centered-cubic ($bcc$) Fe [Jaatinen et al., Physical Review E 80, 031602 (2009)] to demonstrate the procedure as well as to develop physical intuition about the PFC input parameters. We demonstrate that the EOF-PFC model parameterization does not predict stable $bcc$ structures with positive vacancy densities. This result suggests an alternative parameterization of the PFC model, which requires the primary peak position of the two-body direct correlation function to shift as a function of $\bar{n}$. 
\end{abstract}
\maketitle

\section{Introduction}

The phase-field crystal (PFC) model is a simulation approach for studying phenomena that occur on atomic length and diffusive time scales. This is achieved by considering a free energy that is minimized by either a periodic order parameter profile, which represents a solid crystalline phase, or a constant order parameter profile, which represents a liquid phase \cite{Elder:2002ys,Elder:2004vn}. Such a formulation allows the PFC model to describe elastic and plastic deformation, multiple crystal orientations, and free surfaces in non-equilibrium processes \cite{Elder:2004vn}. Consequently, the model has been applied to investigate many important materials phenomena such as dislocation dynamics \cite{Berry:2006p6059,Berry:2008p6063,Stefanovic:2009p607}, nucleation \cite{Granasy:2011te,Toth:2010p6918}, and grain-boundary-energy anisotropy \cite{Elder:2004vn,Berry:2008vg}.

The links between PFC model parameters to measurable quantities in experiments and atomistic simulations were made by Elder et al.\ \cite{Elder:2007p1956} who showed that the PFC model can be derived from the classical density functional theory (cDFT) of freezing \cite{Ramakrishnan:1979ve, Haymet:1981jm}. This derivation provided a statistical mechanical interpretation of the PFC order parameter as an atomic-probability density, which is obtained by taking an ensemble average of the microscopic particle density \cite{Hansen:2006tg}. The derivation also associated the bulk modulus and lattice spacing of a crystal to the curvature and position, respectively, of the first peak of the two-body direct correlation function (DCF), which can be obtained from experiments or atomistic simulations.

Although the PFC model parameters have been linked to measurable quantities, the procedures for calculating equilibrium material properties from the PFC model are not straightforward \cite{Pisutha-Arnond:3001fk} because the thermodynamic interpretation of the PFC free energy has not been fully developed. In this paper, we present a thermodynamic interpretation for bulk phases of the PFC model. As a starting point, we follow the thermodynamic formalism for a crystalline system that was introduced by Larch\'e and Cahn \cite{Larche:1973wp} and was detailed in Voorhees and Johnson \cite{Voorhees:2007tg} to derive a thermodynamic relationship between the PFC free energy and thermodynamic state variables. We then apply the equilibrium conditions between a bulk crystal and liquid from Voorhees and Johnson \cite{Voorhees:2007tg} to the thermodynamic relationship for the PFC model to obtain a thermodynamically consistent procedure for determining solid-liquid phase coexistence, which is demonstrated to be in agreement with the common-tangent construction  commonly used in the PFC community \cite{Elder:2004vn,Jaatinen:2010gp,Greenwood:2011bs}. Finally, we apply this procedure to a PFC model parameterized for body-centered-cubic ($bcc$) Fe  via an eighth-order fit (EOF) of the two-body DCF in Fourier space \cite{Jaatinen:2009p2381}. The EOF-PFC model is used to demonstrate the procedure as well as to examine how the average value of the order parameter, $\bar{n}$, and lattice spacing, $a$, are related to the number of atoms and vacancies in a crystal simulated by the PFC model.

The paper is outlined as follows. In Section \ref{sec:PFC_free_energy}, we review the free energy of the PFC model. In Section \ref{sec:thermodynamics}, we describe the thermodynamics for bulk liquid and crystalline phases. In Section \ref{sec:FED}, we derive the free-energy densities (FED) for the bulk liquid and crystalline phases, which serve as the basis for the thermodynamic interpretation of the PFC free energy. In Section \ref{sec:phase_coexistence}, we present the equilibrium conditions between a bulk crystal and liquid phase and apply these conditions to the thermodynamic relationship for the PFC model to obtain a procedure for determining solid-liquid phase coexistence. In Section \ref{sec:EOF}, this procedure is applied to the EOF-PFC model to demonstrate the procedure, as well as to develop physical intuition about the PFC model parameters. For this model, we derive an upper-bound expression for $\bar{n}$, above which the vacancy density becomes negative. We further show that the EOF-PFC model does not stabilize $bcc$ structures if values of $\bar{n}$ are below the upper bound. These results indicate a need for an alternative parameterization of the PFC model. Finally, in Section \ref{sec:conclusion}, we summarize the results of our work and present potential directions for future work. 

\section{The PFC Free-Energy Functional}
\label{sec:PFC_free_energy}

The PFC model is based on a free-energy difference with respect to a reference liquid phase. The free energy is written in terms of an ideal-gas contribution, $\Delta \mathcal{F}_{id}[n(\textbf{r})]$, and an excess contribution, $\Delta \mathcal{F}_{ex}[n(\textbf{r})]$,
\begin{equation}
\Delta \mathcal{F}[n(\textbf{r})] = \Delta \mathcal{F}_{id}[n(\textbf{r})] + \Delta \mathcal{F}_{ex}[n(\textbf{r})] ,
\label{free_energy}
\end{equation}
where $\Delta \mathcal{F}_{id}[n(\textbf{r})]$ is derived from a system of non-interacting particles and $\Delta \mathcal{F}_{ex}[n(\textbf{r})]$ contains the contribution from the interactions between particles.\cite{Elder:2007p1956} The $\Delta$ symbols denote free-energy differences with respect to a reference liquid phase with (constant) atomic-probability density of $\rho_0$. The scaled dimensionless number density, $n(\textbf{r})$, is related to the atomic-probability density, $\rho(\textbf{r})$, by $n(\textbf{r}) \equiv \rho(\textbf{r})/\rho_0 - 1$. 

Each term in Eq.\ \eqref{free_energy} is written as an integral of a FED, 
\begin{eqnarray}
\Delta \mathcal{F}[n(\textbf{r})] &=& \int_\mathcal{V} \Delta f_{\text{PFC}}(n(\textbf{r})) d\textbf{r} \nonumber\\
 &=& \int_\mathcal{V} \bigg [ \Delta f_{id}(n(\textbf{r})) +  \Delta f_{ex}(n(\textbf{r})) \bigg]d\textbf{r} ,
\label{free_energy_density}
\end{eqnarray}
where $\Delta f_{\text{PFC}}(n(\textbf{r}))$ is the FED of the PFC model, and $\Delta f_{id}(n(\textbf{r}))$ and $\Delta f_{ex}(n(\textbf{r}))$ are the ideal and excess contributions to $\Delta f_{\text{PFC}}(n(\textbf{r}))$, respectively. The subscripts $\mathcal{V}$ denote that the integrals are over the system volume.

The ideal-gas FED,
\begin{equation}
\Delta f_{id}(n(\textbf{r}))  = \rho_0 k_B T \bigg [a_t\frac{n(\textbf{r})^2}{2} - b_t\frac{n(\textbf{r})^3}{6}+\frac{n(\textbf{r})^4}{12} \bigg] ,
\label{free_energy_ideal}
\end{equation}
is obtained by approximating the Helmholtz free energy of an ideal gas with a polynomial expansion. The parameters $a_t$ and $b_t$ are fitting constants, $k_B$ is the Boltzmann constant, and $T$ is the temperature of the reference liquid phase. On the other hand, the excess FED,
\begin{equation}
 \Delta f_{ex}(n(\textbf{r}))=-\frac{\rho_0^2 k_B T}{2}  \int   n(\textbf{r}) C^{(2)}(\mid \textbf{r}-\textbf{r}^{\prime} \mid) n(\textbf{r}^{\prime}) d\textbf{r}^{\prime} ,
\label{free_energy_excess}
\end{equation}
is obtained by expanding the excess Helmholtz free energy to include correlation functions up to the second order, i.e., the two-body DCF, $C^{(2)}$ \cite{Hansen:2006tg}. In writing the two-body DCF as $C^{(2)}(\mid \textbf{r}-\textbf{r}^{\prime} \mid)$, an assumption has been made that the DCF is spherically symmetric.\cite{Elder:2007p1956} Combining Eqs.\ \eqref{free_energy} through \eqref{free_energy_excess}, the PFC free energy is
\begin{equation}
\Delta \mathcal{F}[n(\textbf{r})] = \rho_0 k_B T \int_{\mathcal{V}} \left( a_t\frac{n(\textbf{r})^2}{2} - b_t\frac{n(\textbf{r})^3}{6}+\frac{n(\textbf{r})^4}{12}   -\frac{\rho_0}{2}  \int   n(\textbf{r}) C^{(2)}(\mid \textbf{r}-\textbf{r}^{\prime} \mid) n(\textbf{r}^{\prime}) d\textbf{r}^{\prime} \right) d\textbf{r} .
\label{PFC_free_energy}
\end{equation}
$\Delta \mathcal{F}[n(\textbf{r})]$ is minimized by $n(\textbf{r})$ that is equal to a constant value or that contains peaks with the periodicity of a crystalline lattice. Regions where $n(\textbf{r})$ is constant are considered to be in a liquid state, while those where $n(\textbf{r})$ is periodic are considered to be in a crystalline state.

The calculation of $\Delta f_{ex}(n(\textbf{r}))$ is often performed in Fourier space, 
\begin{equation}
\Delta f_{ex}(n(\textbf{r}))=-\frac{\rho_0^2 k_B T}{2}  n(\textbf{r}) \mathscr{F}^{-1}[\hat{C}^{(2)}(k) \hat{n}(\textbf{k})] ,
\label{free_energy_excess_Fourier}
\end{equation}
where the convolution theorem is used to efficiently evaluate the integral of Eq.\ (\ref{free_energy_excess}) as the inverse Fourier transform of the product of Fourier transforms. The notation $\mathscr{F}^{-1}[ \ ]$ denotes the inverse Fourier transform, $\textbf{k}$ is a wave vector, $k \equiv |\textbf{k}|$, and the hat symbols denote the Fourier transform of the respective quantities. 

Thermodynamics describe the properties of systems that are in equilibrium. In the PFC model, an equilibrium density profile, $n_{eq}(\textbf{r})$, is obtained by relaxing $n(\textbf{r})$ via conserved dissipative dynamics \cite{Elder:2007p1956,Elder:2002ys,Elder:2004vn},
\begin{equation}
\frac{\partial n(\textbf{r})}{ \partial t} = \nabla^2 \frac{\delta \Delta \mathcal{F}[n(\textbf{r})]}{\delta n(\textbf{r})}  ,
\label{dynamics}
\end{equation}
until a stationary state is reached. This state corresponds to the lowest energy state for the given constraints on the order-parameter average, $\bar{n}\equiv (1/\mathcal{V})\int n(\textbf{r})d\textbf{r}$, and lattice spacing, $a$, and will be referred to as the single-phase equilibrium state. In a single-phase equilibrium bulk liquid phase (i.e., away from any interfaces or boundaries), $n_{eq}(\textbf{r}) = n^{\text{bulk},l}_{eq}(\textbf{r}) = \bar{n}$. Thus, a coarse-grained FED for the bulk liquid phase is $\Delta f_{\text{PFC}}^{\text{bulk},l}(\bar{n}) \equiv \Delta f_{id}(\bar{n}) + \Delta f_{ex}(\bar{n})$. On the other hand, in a single-phase equilibrium bulk crystalline phase, $n_{eq}(\textbf{r}) = n^{\text{bulk},c}_{eq}(\textbf{r})$, which has a density profile that is periodic with a uniform amplitude for each value of $\bar{n}$. Therefore, the free energy of a bulk crystalline phase corresponding to $n^{\text{bulk},c}_{eq}(\textbf{r})$ is a function of $\bar{n}$, $a$, and system volume, $\mathcal{V}$. Consequently, the coarse-grained FED of a bulk crystalline phase is given by
\begin{equation}
\Delta f_{\text{PFC}}^{\text{bulk},c}(\bar{n},a) \equiv \lim_{\mathcal{V}\rightarrow\infty} \left (\frac{\Delta \mathcal{F}[n_{eq}^{\text{bulk},c}(\textbf{r})]}{\mathcal{V}} \right) ,
\label{bulk_free_energy_density}
\end{equation}
where the limit indicates that the system volume is large enough such that the bulk phase is far away from any interfaces or boundaries. The definition in Eq.\ \eqref{bulk_free_energy_density} shows that the PFC FED of a bulk crystalline phase is a function of $\bar{n}$ and $a$, and is defined in terms of $\mathcal{V}$. Since $\bar{n}$ and $a$ are coarse-grained variables, a system having $n(\textbf{r}) = n_{eq}^{\text{bulk},c}(\textbf{r})$ for the solid phase and $n(\textbf{r}) = n_{eq}^{\text{bulk},l}(\textbf{r}) = \bar{n}$ for the liquid phase can be described by the thermodynamics of bulk phases.

The value of $a$ that minimizes $\Delta f_{\text{PFC}}^{\text{bulk}}(\bar{n},a)$ for each $\bar{n}$ is denoted as $a^*$. Its value is set by the position of the maximum of the primary (first) peak of the two-body DCF in Fourier space, $k_m$, where $a^* \propto k_m^{-1}$ \cite{Elder:2007p1956,Greenwood:2010uq}. Since the position of the primary peak in the two-body DCF is independent of $\bar{n}$ in the PFC model, the value of $a^*$ remains unchanged for all values of $\bar{n}$. 

The PFC model has several variations that depend on the choice of the two-body DCF. A thorough review of different formulations and extensions of the PFC model is given in Ref.\ \onlinecite{Emmerich:2012ko}. In this paper, we consider the PFC model proposed by Jaatinen et al.\ \cite{Jaatinen:2009p2381} because their model has been parameterized to an experimentally based two-body DCF that was calculated from atomistic simulations of $bcc$ Fe \cite{Wu:2007p1925}. Their PFC model employs an ideal-gas FED with $a_t=0.6917$ and $b_t=0.08540$ and an EOF approximation of the two-body DCF, which has the form
\begin{equation}
\rho_0 \hat{C}^{(2)}_{EOF}(k) = \mathcal{C}_m -\Gamma  \left(\frac{k_m^2-k^2}{k_m^2} \right)^2 
- E_B \left(\frac{k_m^2-k^2}{k_m^2} \right)^4, 
\label{EOF}
\end{equation}
where
\begin{eqnarray}
\Gamma = - \frac{k_m^2 \mathcal{C}_c}{8} , \ \ E_B = \mathcal{C}_m - \mathcal{C}_0 - \Gamma  ,
\end{eqnarray}
and $k_m$, $\mathcal{C}_0$, $\mathcal{C}_m$, and $\mathcal{C}_c$ are fitting constants that are defined in terms of the features of the two-body DCF in Fourier space: $\mathcal{C}_0 \equiv \rho_0\hat{C}^{(2)}_{EOF}(0)$, $\mathcal{C}_m \equiv \rho_0\hat{C}^{(2)}_{EOF}(k_m)$, and $\mathcal{C}_c \equiv d^2 \left(\rho_0\hat{C}^{(2)}_{EOF}(k)\right)/dk^2 |_{k=k_m}$ (curvature at $k=k_m$). In Ref.\ \onlinecite{Jaatinen:2009p2381}, the EOF-PFC model was parameterized to the bulk modulus and solid-liquid interfacial energies of $bcc$ Fe, where the fitting constants have the following values: $k_m=2.985 \AA^{-1}$, $\mathcal{C}_0=-49\AA^{-3}$, $1-\mathcal{C}_m=0.332\AA^{-3}$,  $\mathcal{C}_c=-10.40\AA^{-1}$, and $\rho_0 = 0.0801 \AA^{-3}$. 

\section{Thermodynamics for Bulk Phases}
\label{sec:thermodynamics}

In this section, we review the thermodynamics for bulk liquid and crystalline phases. First, we consider the thermodynamics for a bulk liquid phase, i.e., $n(\textbf{r}) = n^{\text{bulk},l}_{eq}(\textbf{r}) = \bar{n}$, and provide definitions for the entropy, $S^l$, hydrostatic pressure, $P^l$, and chemical potential, $\mu_A^l$, for the bulk liquid phase, where the superscript $l$ denotes quantities associated with the liquid phase. Next, we consider the thermodynamics of a bulk crystalline phase, i.e., $n(\textbf{r}) = n^{\text{bulk},c}_{eq}(\textbf{r})$, which incorporates the description of a lattice with sites that contain either atoms or vacancies, as introduced by Larch\'e and Cahn \cite{Larche:1973wp}. The definitions for the entropy, $S^c$, hydrostatic pressure, $P^c$, and two different chemical potentials, $\mu_A^c$ and $\mu_L^c$, are provided for the bulk crystalline phase, where the superscript $c$ denotes quantities associated with the crystalline phase.   

\subsection{Bulk Liquid Phase}

The free energy of the PFC model is the Helmholtz free energy \cite{Elder:2007p1956}. For the bulk liquid phase, the Helmholtz free energy is
\begin{equation}
F^l = F^l(\theta^l, \mathcal{V}^l, N_A^l),
\label{helmholtz_l}
\end{equation}
where $\theta^l$, $\mathcal{V}^l$, and $N_A^l$ are the temperature, volume, and the number of atoms in the bulk liquid, respectively. The differential form of $F^l$ is
\begin{equation}
dF^l = -S^l d\theta^l - P^l d\mathcal{V}^l + \mu_A^l dN_A^l
\label{dF^l}
\end{equation}
where
\begin{equation}
S^l \equiv -\frac{\partial F^l}{\partial \theta^l} \bigg |_{\mathcal{V}^l, N_A^l}, \ \ \ P^l \equiv -\frac{\partial F^l}{\partial \mathcal{V}^l} \bigg |_{\theta^l, N_A^l}, \ \ \ \mu_A^l \equiv \frac{\partial F^l}{\partial N_A^l} \bigg |_{\theta^l, \mathcal{V}^l}.
\end{equation}
The integrated form of Eq.\ (\ref{dF^l}) is 
\begin{equation}
F^l(\theta^l,\mathcal{V}^l,N_A^l) =  -P^l\mathcal{V}^l + \mu_A^l N_A^l  
\label{F^l}
\end{equation}
and the corresponding Gibbs-Duhem relation is 
\begin{equation}
0 = S^l d\theta^l -\mathcal{V}^ldP^l +N_A^l d\mu_A^l.
\label{Gibbs_Duhem_l}
\end{equation}
Equations\ \eqref{F^l} and \eqref{Gibbs_Duhem_l} will be used later to derive FEDs for the bulk liquid phase.

\subsection{Bulk Crystalline Phase}

For a bulk crystalline solid, a lattice is employed to represent the spatially periodic structure of a crystal, where each lattice site contains either an atom or a vacancy \cite{Larche:1973wp,Voorhees:2007tg}. For a one-component bulk crystalline phase, the total number of lattice sites, $N_L^c$, is related to the number of atoms, $N_A^c$, and vacancies, $N_V^c$, by
\begin{equation}
N_L^c = N_A^c + N_V^c.
\label{N_L}
\end{equation}
As discussed in Voorhees and Johnson \cite{Voorhees:2007tg}, any two of the three thermodynamic variables in Eq.\ \eqref{N_L} can be used to describe the thermodynamic state of a one-component crystal. In this work we consider the Helmholtz free energy as a function of $N_A^c$ and $N_L^c$, 
\begin{equation}
F^c = F^c(\theta^c,\mathcal{V}^c,N_A^c,N_L^c) ,
\label{helmholtz_c}
\end{equation}
where $\theta^c$ and $\mathcal{V}^c$ are the temperature and the crystal volume, respectively. The differential form of $F^c$ is
\begin{equation}
dF^c = -S^c d\theta^c -P^c d\mathcal{V}^c + \mu_A^c dN_A^c + \mu_L^c dN_L^c
\label{dF^c}
\end{equation}
where
\begin{equation}
S^c \equiv -\frac{\partial F^c}{\partial \theta^c} \bigg |_{\mathcal{V}^c, N_A^c, N_L^c}, \ \ \ P^c \equiv -\frac{\partial F^c}{\partial \mathcal{V}^c} \bigg |_{\theta^c, N_A^c, N_L^c}, \ \ \ \mu_A^c \equiv \frac{\partial F^c}{\partial N_A^c} \bigg |_{\theta^c, \mathcal{V}^c, N_L^c}, \ \ \ \mu_L^c \equiv \frac{\partial F^c}{\partial N_L^c} \bigg |_{\theta^c, \mathcal{V}^c, N_A^c} .
\label{partial_derivatives}
\end{equation}
The partial derivative that defines $\mu_A^c$ provides the energy change due to the addition or removal of an atom while the number of lattice sites and crystal volume are held constant for an isothermal system. The definition of $\mu_A^c$ requires a vacancy to be eliminated when an atom is added to the crystal and a vacancy to be generated when an atom is removed from the crystal. This chemical potential is hereafter referred to as the diffusion potential \cite{Voorhees:2007tg} to distinguish from the chemical potential of atoms for the bulk liquid phase. On the other hand, the partial derivative that defines $\mu_L^c$ provides the energy change due to a change in the number of lattice sites while the crystal volume and the number of atoms are held constant for an isothermal system. A lattice site can be added by moving an atom within the crystal to the surface while simultaneously creating a vacancy within the crystal. This process will cause an increase in the pressure when the crystal volume is held constant and the partial molar volume of the vacancy is nonzero \cite{Voorhees:2007tg}.

The integrated form of Eq.\ (\ref{dF^c}) is 
\begin{equation}
F^c(\theta^c,\mathcal{V}^c,N_A^c,N_L^c) =  -P^c\mathcal{V}^c + \mu_A^c N_A^c + \mu_L^c N_L^c  
\label{F^c}
\end{equation}
and the corresponding Gibbs-Duhem relation is 
\begin{equation}
0 = S^c d\theta^c -\mathcal{V}^cdP^c +N_A^c d\mu_A^c + N_L^c d\mu_L^c,
\label{Gibbs_Duhem_c}
\end{equation}
the derivation of which is presented in appendix \ref{Appendix:Gibbs_Duhem}. Equations (\ref{F^c}) and (\ref{Gibbs_Duhem_c}) will be used later to derive FEDs for a bulk crystalline phase. 

\section{Free-Energy Densities for Bulk Phases}
\label{sec:FED}

In this section, we derive FEDs that are defined on a reference (undeformed) volume, $\mathcal{V}^{\prime}$, and the (potentially deformed) system volumes, $\mathcal{V}^l$ and $\mathcal{V}^c$, for the bulk liquid and crystalline phases, respectively. FEDs defined on $\mathcal{V}^{\prime}$ are referred to as reference-volume FEDs, while those defined on  $\mathcal{V}^l$ or $\mathcal{V}^c$ are referred to as system-volume FEDs. By defining the reference-volume FED for the bulk crystalline phase, we make an important distinction between two sources of pressure change: mechanical and configurational forces \cite{Voorhees:2007tg}. The system-volume FEDs are used to develop a thermodynamic relationship between the PFC FED and thermodynamic state variables. 

\subsection{Bulk Liquid Phase}

The reference-volume FED for the bulk liquid phase, denoted as $f^l_{\mathcal{V}^{\prime}}$, is obtained by dividing $F^l$ by $\mathcal{V}^{\prime}$,
\begin{equation}
f^{l}_{\mathcal{V}^{\prime}} \equiv \frac{F^l}{\mathcal{V}^{\prime}} = -P^l J^l + \mu_A^l \rho^{\prime l}_{A} ,
\label{f^l_rV}
\end{equation}
where 
\begin{equation}
J^l \equiv \frac{\mathcal{V}^l}{\mathcal{V}^{\prime}}, \ \ \ \rho^{\prime l}_{A} \equiv \frac{N_A^l}{\mathcal{V}^{\prime}} .
\label{J^l}
\end{equation}
The variable $J^l$ describes volume change due to hydrostatic pressure and $\rho^{{\prime} l}_{A}$ is the atomic density of the liquid phase defined on $\mathcal{V}^{\prime}$, as  denoted by the prime symbol. Similarly, the Gibbs-Duhem relation in Eq.\ (\ref{Gibbs_Duhem_l}) is divided by $\mathcal{V}^{\prime}$ to obtain
\begin{equation}
0 = s^l_{\mathcal{V}^{\prime}} d\theta^l -J^l dP^l + \rho^{{\prime}l}_{A} d\mu^l_A  ,
\label{Gibbs_Duhem_f^l_rV}
\end{equation}
where $s^l_{\mathcal{V}^{\prime}} \equiv S^l/\mathcal{V}^{\prime}$. Differentiating Eq.\ (\ref{f^l_rV}) and subtracting Eq.\ (\ref{Gibbs_Duhem_f^l_rV}) gives an expression for $df^l_{\mathcal{V}^{\prime}}$,
\begin{equation}
df^l_{\mathcal{V}^{\prime}} = - s^l_{\mathcal{V}^{\prime}} d\theta^l - PdJ^l + \mu_A^l d\rho^{\prime l}_{A}  ,
\label{df_p}
\end{equation}
where $f^l_{\mathcal{V}^{\prime}}$ is a function of natural variables $\theta^l$, $J^l$, and $\rho^{\prime l}_{A}$. 

Alternatively, the system-volume FED for the bulk liquid phase, as denoted by $f^l_{\mathcal{V}}$, is obtained by dividing $F^l$ by $\mathcal{V}^l$,
\begin{equation}
f^l_{\mathcal{V}} \equiv \frac{F^l}{\mathcal{V}^l} = -P^l  + \mu_A^l \rho_A^l ,
\label{f^l_sV}
\end{equation}
where $\rho_A^l \equiv N_A^l/\mathcal{V}^l$. Similarly, the Gibbs-Duhem relation in Eq.\ \eqref{Gibbs_Duhem_l} is divided by $\mathcal{V}^l$ to obtain
\begin{equation}
0 = s^l_{\mathcal{V}} d\theta^l - dP^l + \rho_A^l d\mu_A^l,
\label{Gibbs_Duhem_f^l_sV}
\end{equation}
where $s^l_{\mathcal{V}} \equiv S^l/\mathcal{V}^l$. The variables $s^l_{\mathcal{V}}$ and $\rho_A^l$ are defined on the system volume of the bulk liquid phase, $\mathcal{V}^l$. Differentiating Eq.\ (\ref{f^l_sV}) and subtracting Eq.\ (\ref{Gibbs_Duhem_f^l_sV}) gives a relationship for $df^l_{\mathcal{V}}$,
\begin{equation}
df^l_{\mathcal{V}} = -s^l_{\mathcal{V}} d\theta^l +\mu_A^l d\rho_A^l,
\label{df^l_sV}
\end{equation}
where $f^l_{\mathcal{V}}$ is a function of natural variables $\theta^l$ and $\rho_A^l$. Equation\ \eqref{df^l_sV} is defined in terms of the system volume of the liquid phase and is a function of $\rho_A^l$, which is related to $\bar{n}$ of the PFC FED for a bulk liquid via $\bar{n} = \rho_A^l/\rho_0 -1$. Therefore, Eq.\ \eqref{df^l_sV} will be employed to develop a thermodynamic interpretation of the PFC free energy for the bulk liquid phase.

\subsection{Bulk Crystalline Phase}

The reference-volume FED for the bulk crystalline phase, denoted as $f^{c}_{\mathcal{V}^{\prime}}$, is determined by dividing $F^c$ with $\mathcal{V}^{\prime}$,
\begin{equation}
f^{c}_{\mathcal{V}^{\prime}} \equiv \frac{F^c}{\mathcal{V}^{\prime}} = -P^c J^c + \mu_A^c \rho^{\prime c}_{A} + \mu_L^c \rho^{\prime c}_{L}  ,
\label{f^c_rV}
\end{equation}
where 
\begin{equation}
J^c \equiv \frac{\mathcal{V}^c}{\mathcal{V}^{\prime}}, \ \ \ \rho^{\prime c}_{A} \equiv \frac{N_A^c}{\mathcal{V}^{\prime}}, \ \ \ \rho^{\prime c}_L \equiv \frac{N_L^c}{\mathcal{V}^{\prime}}.
\label{J^c}
\end{equation}
Similar to the bulk liquid phase, $J^c$ describes volume change due to hydrostatic pressure in a bulk crystalline phase, and $\rho^{\prime c}_{A}$ and $\rho^{\prime c}_L$ are the atomic and lattice densities, respectively, that are defined on the reference volume. The Gibbs-Duhem relation in Eq.\ (\ref{Gibbs_Duhem_c}) is divided by $\mathcal{V}^{\prime}$ to obtain
\begin{equation}
0 = s^c_{\mathcal{V}^{\prime}} d\theta^c -J^c dP^c + \rho^{\prime c}_{A} d\mu_A^c + \rho^{\prime c}_L d\mu_L^c ,
\label{Gibbs_Duhem_f^c_rV}
\end{equation}
where $s^c_{\mathcal{V}^{\prime}} \equiv S^c/\mathcal{V}^{\prime}$. Differentiating Eq.\ (\ref{f^c_rV}) and subtracting Eq.\ (\ref{Gibbs_Duhem_f^c_rV}) gives an expression for $df^c_{\mathcal{V}^{\prime}}$,
\begin{equation}
df^c_{\mathcal{V}^{\prime}} = - s^c_{\mathcal{V}^{\prime}} d\theta^c - P^cdJ^c + \mu_A^c d\rho^{\prime c}_{A} + \mu^c_L d\rho^{\prime c}_L ,
\label{df^c_rV}
\end{equation}
where $f^c_{\mathcal{V}^{\prime}}$ is a function of natural variables $\theta^c$, $J^c$, $\rho^{\prime c}_{A}$, and $\rho^{\prime c}_L$. 

Changing the volume of an isothermal bulk crystalline system (reflected by a change in $J^c$ because $\mathcal{V}^{\prime}$ is constant) while keeping the mass  (equivalent to fixing $\rho^{\prime c}_{A})$ and number of lattice sites (equivalent to fixing $\rho^{\prime c}_{L})$ constant will cause the pressure to change. This pressure change arises from deforming the system with a \textit{mechanical force}. Alternatively, as mentioned earlier, a pressure change also arises when the number of lattice sites change (reflected by a change in $\rho^{\prime c}_L$) while the crystal volume and mass are constant. To understand this latter type of pressure change, consider a thought experiment where a crystal  is enclosed by a rigid wall that does not allow mass transfer (fixed $\mathcal{V}^c$ and $\rho_A^{\prime c}$). When a lattice site (vacancy) is added to the system, the constraint imposed by the walls prevent a volume change, and thus results in a pressure change. This type of pressure change arises from a \textit{configurational force} \cite{Voorhees:2007tg}. Therefore, the thermodynamic framework for crystalline solids described above allows us to distinguish between pressure changes due to mechanical and configurational forces \footnote{The thermodynamic framework for a crystal has also been used to distinguish between volume change due to configurational and mechanical forces \cite{Voorhees:2007tg}.}. 

As mentioned earlier, the PFC FED is defined on the system volume. Therefore, a system-volume FED for a bulk crystalline phase will be most appropriate for developing a thermodynamic relationship for the PFC model of a bulk crystal. To reformulate the reference-volume FED, Eq.\ \eqref{f^c_rV}, and Eq.\ \eqref{Gibbs_Duhem_f^c_rV} in terms of the system volume, we divide both sides of the equations by $J^c$ (see Eq.\ \eqref{J^c} for definition):
\begin{equation}
\mathbbm{f}^c_{\mathcal{V}} \equiv \frac{f^c_{\mathcal{V}^{\prime}}}{J^c} = -P^c  + \mu_A^c \frac{\rho^{\prime c}_{A}}{J^c} + \mu^c_L \frac{\rho^{\prime c}_L}{J^c} 
\label{fbbm}
\end{equation}
and
\begin{equation}
0 = \frac{s^c_{\mathcal{V}^{\prime}}}{J^c} d\theta^c - dP^c + \frac{\rho^{\prime c}_{A}}{J^c} d\mu^c_A + \frac{\rho^{\prime c}_L}{J^c} d\mu^c_L   ,
\label{Gibbs_Duhem_fbbm}
\end{equation}
where $\mathbbm{f}^c_{\mathcal{V}}$ is one expression for the system-volume FED. The division by $J^c$ above maps $\rho^{\prime c}_{A}$, $\rho^{\prime c}_L$, and $s^c_{\mathcal{V}^{\prime}}$ to their system-volume counterparts: $\rho^c_A$, $\rho^c_L$, and $s^c_{\mathcal{V}}$, respectively. The differential form of Eq.\ \eqref{fbbm} is obtained by taking its derivative and subtracting Eq.\ \eqref{Gibbs_Duhem_fbbm},
\begin{equation}
d\mathbbm{f}^c_{\mathcal{V}} = -\frac{s^c_{\mathcal{V}^{\prime}}}{J^c} d\theta^c +\frac{\mu^c_A}{J^c} d\rho^{\prime c}_{A} + \frac{\mu^c_L}{J^c}d\rho^{\prime c}_{L} + \left(\mu^c_A \rho^{\prime c}_{A} + \mu^c_L\rho^{\prime c}_{L}\right)d(1/J^c)  ,
\label{df_bbm}
\end{equation}
where it can now be observed that $\mathbbm{f}^c_{\mathcal{V}}$ is a function of natural variables $\theta^c$, $\rho^{\prime c}_{A}$, $\rho^{\prime c}_L$, and $1/J^c$. Therefore, the FED denoted by $\mathbbm{f}^c_{\mathcal{V}}$ is a system-volume FED that is a function of densities defined on the reference volume. 

In order to obtain a FED with independent variables that match those of the PFC FED, the variables $\rho^{\prime c}_{A}$ and $\rho^{\prime c}_L$ are related to $\rho^c_A$ and $\rho^c_L$ by the chain rule
\begin{equation}
d\rho^{\prime c}_{A} = d(\rho^c_A J^c) = J^c d\rho^c_A + \rho^c_A dJ^c \ \ \text{ and } \ \  d\rho^{\prime c}_L = d(\rho^c_L J^c) = J^c d\rho^c_L + \rho^c_L dJ^c  .
\label{drhop_to_drho}
\end{equation}
Substituting Eq.\ \eqref{drhop_to_drho} into Eq.\ \eqref{df_bbm}, one obtains
\begin{equation}
df^c_{\mathcal{V}} = -s^c_{\mathcal{V}} d\theta^c + \mu^c_A d\rho^c_A + \mu^c_L d\rho^c_L  ,
\label{df^c_sV}
\end{equation}
%
where we have used the relationships
\begin{equation}
d J^c = -(J^c)^2 d(1/J^c) \ \ \text{ and } \ \ s^c_{\mathcal{V}} = \frac{s^c_{\mathcal{V}^{\prime}}}{J^c} .
\label{J_relation}
\end{equation}
The same expression can be obtained by starting from a FED based on the system volume; however, we derive it in this manner to explicitly illustrate the connection between reference-volume and system-volume variables. This expression of the system-volume FED for the bulk crystal, $f^c_{\mathcal{V}}$, given in Eq.\ \eqref{df^c_sV} is now a function of natural variables $\theta^c$, $\rho^c_A$, and $\rho^c_L$. By expressing $\rho^c_L$ in terms of the volume of a lattice site, $\mathcal{V}_L^c$,
\begin{equation}
\rho^c_L  = \frac{1}{\mathcal{V}^c_L}  ,
\label{rhoL}
\end{equation}
Eq.\ \eqref{df^c_sV} is related to the lattice spacing via the unit-cell volume, $\mathcal{V}^c_C$, which is written in terms of $\mathcal{V}_L^c$ using the number of lattice sites per unit cell, $\chi^c \equiv \mathcal{V}^c_C/\mathcal{V}^c_L$. The value of $\chi^c$ depends on the lattice structure. For example, a $bcc$ structure, which contains 2 lattice sites per unit cell, has $\chi^c = 2$. 

Equation (\ref{rhoL}) is used to express $df^c_{\mathcal{V}}$ as
\begin{equation}
df^c_{\mathcal{V}} = -s^c_{\mathcal{V}} d\theta^c +  \mu^c_A d\rho^c_A + \chi^c \mu^c_Ld(1/\mathcal{V}^c_C) ,
\label{df^c_uV}
\end{equation}
%
where $f^c_{\mathcal{V}}$ is a function of natural variables $\theta^c$, $\rho^c_A$, and $1/\mathcal{V}^c_C$. Since $\mathcal{V}^c_C$ is a sole function of $a$ (e.g., for a $bcc$ structure, $\mathcal{V}^c_C = a^3$), the FED representation in Eq.\ \eqref{df^c_uV} is a function of natural variables that correspond to those of the PFC FED of the bulk crystal. Therefore, Eq.\ (\ref{df^c_uV}) is employed in developing a thermodynamic interpretation of the PFC free energy for a bulk crystalline phase. The integrated form of Eq.\ (\ref{df^c_uV}) is derived from Eqs.\ (\ref{fbbm}) and (\ref{rhoL}) to be
\begin{equation}
f^c_{\mathcal{V}} = -P^c + \mu^c_{A} \rho^c_A + \frac{\chi^c \mu^c_L}{\mathcal{V}^c_C} ,
\label{f^c_uV}
\end{equation}
%
and Eqs.\ \eqref{drhop_to_drho} and \eqref{rhoL} are substituted into Eq.\ \eqref{Gibbs_Duhem_fbbm} to obtain
\begin{equation}
0 = -s^c_{\mathcal{V}} d\theta^c - dP^c + \rho^c_A d\mu^c_A + \frac{\chi^c}{\mathcal{V}^c_C} d\mu^c_L  .
\label{Gibbs_Duhem_f^c_uV}
\end{equation}
The expressions in Eqs.\ (\ref{df^c_uV}) and (\ref{f^c_uV}) form the basis for deriving a thermodynamic relationship for a bulk crystalline phase in the PFC model.

As observed from Eq.\ \eqref{df^c_uV}, $\mu^c_A = \partial f^c_{\mathcal{V}}/\partial \rho^c_A|_{\theta^c,\mathcal{V}^c_C}$ and thus changing $\rho^c_A$ while holding $\mathcal{V}^c_C$ and $\theta^c$ constant is equivalent to changing the number of atoms while the number of lattice sites and crystal volume are held constant (see Eq.\ \eqref{partial_derivatives}). A negative value of $\mu^c_A$ indicates the presence of a driving force for adding an atom into a vacant lattice site. On the other hand, a positive value of $\mu^c_A$ indicates the presence of a driving force for removing an atom from an occupied lattice site.  Additionally, it can be observed that $\chi^c \mu^c_L = \partial f^c_{\mathcal{V}}/\partial (1/\mathcal{V}^c_C)|_{\theta^c,\rho_A^c}$ and thus changing $\mathcal{V}^c_C$ (equivalent to changing lattice spacing) while holding $\rho_A^c$ and $\theta^c$ constant is equivalent to changing the number of lattice sites while the crystal volume and mass are held constant. Therefore, the process of changing $\mathcal{V}^c_C$ while holding $\rho^c_A$ constant gives rise to a configurational force, just as in the case for changing the number of lattice sites as indicated in Eq.\ \eqref{partial_derivatives}. This point will be further examined in the next section after the FED in Eq.\ \eqref{df^c_uV} is linked to the PFC FED.

\section{Isothermal Thermodynamic Relationships for the Phase-Field Crystal Model}

In this section, we use the system-volume FEDs derived in the previous section to develop the relationship of the PFC FED to thermodynamic state variables for the bulk liquid and crystalline phases. Here we limit the scope to isothermal systems, which correspond to a fixed two-body DCF in the PFC model. Therefore, $d\theta=0$ for our formulation below.

\subsection{Bulk Liquid Phase}

Since the PFC model is based on a free-energy difference from a reference liquid phase, the PFC FED of a bulk liquid phase is related to $f^l_{\mathcal{V}}$ in Eq.\ (\ref{f^l_sV}) by 
\begin{equation}
\Delta f_{\text{PFC}}^{\text{bulk},l}(\bar{n}) = f^l_{\mathcal{V}} - f_0,
\label{f_PFC_to_f^l_sV}
\end{equation}
where $f_0$ is the Helmholtz FED of the reference liquid phase with a density of $\rho_0$. Note that $f_0$ remains constant as $\bar{n}$ changes because the two-body DCF is taken at the reference state and is assumed to be independent of $\bar{n}$. Furthermore, $\bar{n}$ is related to $\rho^l_A$ of Eq.\ (\ref{f^l_sV}) by
\begin{equation}
\rho^l_A = \bar{n} \rho_0 + \rho_0 .
\label{rho^l_A_to_n_bar}
\end{equation}
Equations (\ref{f_PFC_to_f^l_sV}) and \eqref{rho^l_A_to_n_bar} are combined with Eq.\ (\ref{f^l_sV}) to obtain
\begin{equation}
\Delta f_{\text{PFC}}^{\text{bulk},l}(\bar{n}) \equiv f^l_{\mathcal{V}} - f_0 =  -P^l + \rho_0 \mu^l_A \left(\bar{n} + 1 \right) - f_0 .
\label{f^l_PFC}
\end{equation}
Similarly, Eq.\ \eqref{rho^l_A_to_n_bar} is combined with Eq.\ \eqref{Gibbs_Duhem_f^l_sV} to obtain
\begin{equation}
0 =  - dP^l + \rho_0(\bar{n}+1) d\mu_A^l.
\label{Gibbs_Duhem_f^l_PFC}
\end{equation}

Differentiating Eq.\ (\ref{f^l_PFC}) and subtracting Eq.\ \eqref{Gibbs_Duhem_f^l_PFC} gives an expression for $d (\Delta f_{\text{PFC}}^{\text{bulk},l}(\bar{n}))$,
\begin{equation}
d ( \Delta f_{\text{PFC}}^{\text{bulk},l}(\bar{n}) ) \equiv d\left( f^l_{\mathcal{V}} - f_0 \right) = \rho_0 \mu^l_Ad \bar{n} .
\label{df^l_PFC}
\end{equation}

\subsection{Bulk Crystalline Phase}

The PFC FED of a bulk crystalline phase is related to $f^c_{\mathcal{V}}$ in Eq.\ (\ref{f^c_uV}) by 
\begin{equation}
\Delta f_{\text{PFC}}^{\text{bulk},c}(\bar{n},a) = f^c_{\mathcal{V}} - f_0,
\label{f_PFC_to_f^c_sV}
\end{equation}
and $\bar{n}$ is related to $\rho^c_A$ of Eq.\ (\ref{f^c_uV}) by
\begin{equation}
\rho^c_A = \bar{n} \rho_0 + \rho_0 .
\label{rho^c_A_to_n_bar}
\end{equation}
Equations (\ref{f_PFC_to_f^c_sV}) and \eqref{rho^c_A_to_n_bar} are combined with Eq.\ (\ref{f^c_uV}) to obtain 
\begin{equation}
\Delta f_{\text{PFC}}^{\text{bulk},c}(\bar{n},a) \equiv f^c_{\mathcal{V}} - f_0 =  -P^c + \rho_0 \mu^c_A \left(\bar{n} + 1 \right) + \frac{\chi^c \mu^c_L}{\mathcal{V}^c_C}  - f_0  .
\label{f^c_PFC}
\end{equation}
Similarly, Eq.\ \eqref{rho^c_A_to_n_bar} is combined with Eq.\ \eqref{Gibbs_Duhem_f^c_uV} to obtain
\begin{equation}
0 = - dP^c + \rho_0 (\bar{n}+1) d\mu^c_A + \frac{\chi^c}{\mathcal{V}^c_C} d\mu^c_L  .
\label{Gibbs_Duhem_f^c_PFC}
\end{equation}

Differentiating Eq.\ (\ref{f^c_PFC}) and subtracting Eq.\ \eqref{Gibbs_Duhem_f^c_PFC} gives an expression for $d (\Delta f_{\text{PFC}}^{\text{bulk},c}(\bar{n},a))$,
\begin{equation}
d ( \Delta f_{\text{PFC}}^{\text{bulk},c}(\bar{n},a) ) \equiv d\left( f^c_{\mathcal{V}} - f_0 \right) = \rho_0\mu^c_Ad \bar{n}  + \chi^c \mu^c_L d(1/\mathcal{V}^c_C),
\label{df^c_PFC}
\end{equation}
where $\mathcal{V}^c_C$ is a function of $a$. 

Equation \eqref{df^c_PFC} is central to this work because it links the PFC FED to $\mu_A^c$ and $\mu_V^c$, which are chemical potentials that correspond to different thermodynamic processes. As seen in Eq.\ \eqref{df^c_PFC}, $\mu_A^c = 1/\rho_0(\partial \Delta f_{\text{PFC}}^{\text{bulk},c}/\partial \bar{n} |_{\mathcal{V}^c_C})$ and therefore varying $\bar{n}$ when the temperature and lattice spacing are held constant in the PFC model is equivalent to changing the number of atoms while $\theta^c$, $\mathcal{V}^c$, and $N^c_L$ are fixed. Similarly, since $\mu_L^c = 1/\chi^c (\partial \Delta f_{\text{PFC}}^{\text{bulk},c}/\partial (1/\mathcal{V}^c_C) |_{\bar{n}}$), varying lattice spacing when temperature and $\bar{n}$ are held constant in the PFC model is equivalent to changing the number of lattice sites while $\theta^c$, $\mathcal{V}^c$, and $N^c_A$ are fixed.

As discussed earlier, the addition or removal of lattice sites while the crystal volume is held constant gives rise to a pressure change due to a configurational force. This point has not previously been elucidated, and the resulting pressure has been instead attributed to pressure change due to mechanical forces, leading to an improper procedure for elastic constant calculations with the PFC model \cite{Elder:2004vn, Wu:2010uq, Greenwood:2011bs}. A thermodynamically consistent procedure for calculating elastic constants was developed in our previous work \cite{Pisutha-Arnond:3001fk}, and the framework above further validates our approach. 


\section{Solid-Liquid Phase Coexistence in the PFC Model}
\label{sec:phase_coexistence}

In this section, we apply the thermodynamic relationship developed in the previous section to derive a procedure for determining solid-liquid phase coexistence in the PFC model. First, we describe the equilibrium conditions presented in Voorhees and Johnson \cite{Voorhees:2007tg} between bulk crystalline and liquid phases, which involve constraints on $P$, $\mu_A$, and $\mu_L^c$. These equilibrium conditions are then imposed on the thermodynamic relationship for the PFC model, Eqs.\ \eqref{df^l_PFC} and \eqref{df^c_PFC}, to obtain a thermodynamically consistent procedure for determining solid-liquid phase coexistence.

\subsection{Solid-Liquid Phase Coexistence in the Phase-Field Crystal Model}

As described in Voorhees and Johnson \cite{Voorhees:2007tg}, a bulk crystalline and liquid phase are in equilibrium when the variation in the total energy of a system vanishes. They showed that the variation in the system energy vanishes when thermal, mechanical, and chemical equilibria are achieved. For a one-component bulk crystalline phase in equilibrium with a bulk liquid phase, the equilibrium conditions are
\begin{equation}
\theta^c = \theta^l, \ \ \ P^c=P^l, \ \ \ \mu_A^c = \mu_A^l, \ \ \ \mu_L^c=0 .
\label{sl_criteria}
\end{equation}
The last condition of $\mu_L^c=0$ is unique to crystalline solids and indicates that there is no driving force for an addition or removal of a lattice site. This condition can be used to calculate equilibrium vacancy concentration for a crystal in equilibrium with a liquid phase \cite{Voorhees:2007tg}. 

The condition for thermal equilibrium is met when considering an isothermal system. Furthermore, $\mu^c_L=0$ when the FED of the bulk crystalline phase, $\Delta f_{\text{PFC}}^{\text{bulk},c}(\bar{n},a)$, is relaxed with respect to the lattice spacing while $\bar{n}$ and $\theta^c$ are held constant (see Eq.\ \eqref{df^c_PFC}). Therefore, the task of finding solid-liquid coexistence lies in satisfying the equilibrium conditions, $P^c = P^l$ and $\mu^c_A = \mu^l_A$.

\begin{figure}[h]
\begin{center}
$
\begin{array}{c}
\includegraphics[height=7cm]{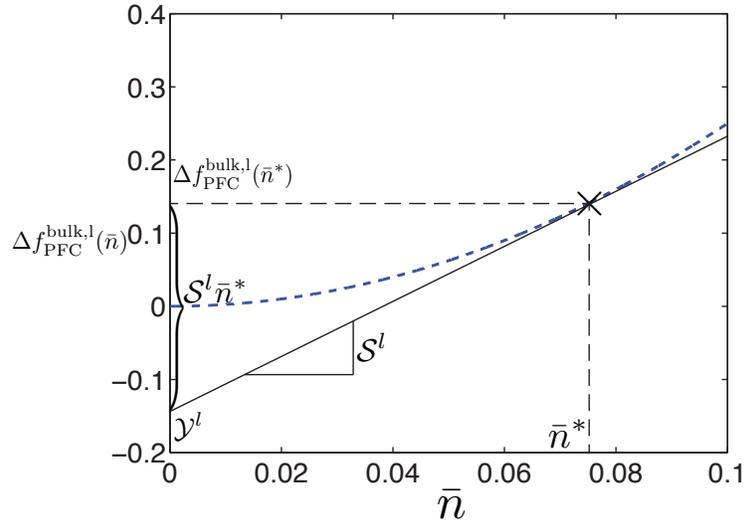} 
\end{array}
$
\end{center}
\caption{\label{fig:Y_and_S_schematic} Schematic of Eq.\ \eqref{Y^l_S^l} on a liquid FED curve (dashed line) with tangent line (solid line) at the point $\bar{n} = \bar{n}^*$, which is marked with ``$\times$". The schematic for Eq.\ \eqref{Y^c_S^c} is similar, but for the FED curve of a bulk crystalline phase.}
\end{figure}

A relationship for pressure is obtained by rearranging the expressions for the PFC FEDs in Eqs.\ \eqref{f^l_PFC} and \eqref{f^c_PFC}:
\begin{equation}
-P^l =  \Delta f_{\text{PFC}}^{\text{bulk},l}(\bar{n}) - \rho_0 \mu^l_A \left(\bar{n} + 1 \right) +f_0  \ \ \text{and} \ \ 
 -P^c = \Delta f_{\text{PFC}}^{\text{bulk},c}(\bar{n},a) - \rho_0 \mu^c_A \left(\bar{n} + 1 \right)-\frac{\chi^c \mu^c_L}{\mathcal{V}^c_C} +f_0,
\label{P1}
\end{equation}
respectively, where the determination of  $P^l$ and $P^c$ require a knowledge of $f_0$. However, since both the liquid and solid phases have the same reference state, $f_0$ will cancel when equating their pressures. By using
\begin{equation}
\rho_0 \mu_A^l = \frac{\partial \Delta f_{\text{PFC}}^{\text{bulk},l}}{\partial \bar{n}} \bigg |_{\theta^l}
\label{mu_A^l}
\end{equation}
obtained from Eq.\ \eqref{df^l_PFC}, $P^l$ in terms of $\Delta f_{\text{PFC}}^{\text{bulk},l}(\bar{n})$ and $\bar{n}$ is expressed as 
\begin{eqnarray}
-P^l &=&  \Delta f_{\text{PFC}}^{\text{bulk},l}(\bar{n})  -  \frac{\partial \Delta f_{\text{PFC}}^{\text{bulk},l}}{\partial \bar{n}}\bigg |_{\theta^l} (\bar{n}+1) + f_0 \nonumber\\
&=& \mathcal{Y}^l  - \mathcal{S}^l + f_0 
\label{P3}
\end{eqnarray}
where 
\begin{equation}
\mathcal{Y}^l \equiv \Delta f_{\text{PFC}}^{\text{bulk},l}(\bar{n})  -  \mathcal{S}^l \bar{n} \ \ \ \text{and} \ \ \ \mathcal{S}^l \equiv \frac{\partial \Delta f_{\text{PFC}}^{\text{bulk},l}}{\partial \bar{n}}\bigg |_{\theta^l}
\label{Y^l_S^l}
\end{equation}
are the y-intercept and slope of a $\Delta f_{\text{PFC}}^{\text{bulk},l}(\bar{n})$ vs.\ $\bar{n}$ curve, respectively, when $\theta^l$ is constant. A schematic of Eq.\ \eqref{Y^l_S^l} is shown in Fig.\ \ref{fig:Y_and_S_schematic}.
%

Similarly, by using 
\begin{equation}
\rho_0 \mu_A^c = \frac{\partial \Delta f_{\text{PFC}}^{\text{bulk},c}}{\partial \bar{n}} \bigg |_{\theta^c,\mu^c_L}
\label{mu_A^c}
\end{equation}
obtained from Eq.\ \eqref{df^c_PFC}, $P^c$ in terms of $\Delta f_{\text{PFC}}^{\text{bulk},c}(\bar{n},a)$ and $\bar{n}$ is expressed as
\begin{eqnarray}
-P^c &=&\Delta f_{\text{PFC}}^{\text{bulk},c}(\bar{n},a) -  \frac{\partial \Delta f_{\text{PFC}}^{\text{bulk},c}}{\partial \bar{n}}\bigg |_{\theta^c,\mu^c_L} \left( \bar{n} +1 \right) -\frac{\chi^c \mu^c_L}{\mathcal{V}^c_C}+ f_0\nonumber\\ 
&=& \mathcal{Y}^c  - \mathcal{S}^c-\frac{\chi^c \mu^c_L}{\mathcal{V}^c_C}+ f_0 ,
\label{P2}
\end{eqnarray}
where 
\begin{equation}
\mathcal{Y}^c \equiv  \Delta f_{\text{PFC}}^{\text{bulk},c}(\bar{n},a)  -  \mathcal{S}^c \bar{n}, \ \ \ \text{and} \ \ \ \mathcal{S}^c \equiv \frac{\partial \Delta f_{\text{PFC}}^{\text{bulk},c}}{\partial \bar{n}}\bigg |_{\theta^c,\mu^c_L}
\label{Y^c_S^c}
\end{equation}
are the y-intercept and slope of a $\Delta f_{\text{PFC}}^{\text{bulk},c}$ vs.\ $\bar{n}$ curve, respectively, when $\mu^c_L$ and $\theta^c$ are constant. Note that the condition $\mu^c_L=0$ has not been imposed in Eq.\ \eqref{P2}, but will be applied later to fulfill the equilibrium conditions listed in Eq.\ \eqref{sl_criteria}. 

According to Eqs.\ (\ref{P3}) and \eqref{P2}, mechanical equilibrium for solid-liquid phase coexistence is fulfilled when $\mathcal{Y}^c = \mathcal{Y}^l$, $\mathcal{S}^c = \mathcal{S}^l$, and $\mu^c_L=0$. Furthermore, the condition, $\mu^c_A = \mu^l_A$, for chemical equilibrium is also satisfied when $\mathcal{S}^c = \mathcal{S}^l$. Therefore, satisfying the common-tangent condition (i.e., the y-intercept and slope of the two curves are equal) for the $\Delta f_{\text{PFC}}^{\text{bulk},c}$ vs.\ $\bar{n}$ and $f_{\text{PFC}}^{\text{bulk},l}(\bar{n})$ vs.\ $\bar{n}$ curves with $\mu^c_L=0$ fulfills the equilibrium conditions for solid-liquid phase coexistence. This procedure, which involves thermal, mechanical, and chemical equilibria, is in agreement with the common-tangent construction commonly used for calculating solid-liquid phase coexistence in the PFC model \cite{Elder:2004vn,Jaatinen:2010gp, Greenwood:2011bs}. Therefore, the above analysis justifies the use of the common-tangent construction to determine solid-liquid phase coexistence in the PFC model.

As noted earlier, the condition of $\mu^c_L=0$ is accomplished in the PFC model by minimizing the PFC FED for the bulk crystalline phase with respect to $\mathcal{V}^c_C$ while $\theta^c$ and $\bar{n}$ are held constant, as seen in Eq.\ \eqref{df^c_PFC}. Therefore, the minimization of the density profile according to the PFC FED  with respect to lattice spacing for a crystalline solid, which is conventionally done in the PFC model \cite{Jaatinen:2010gp}, is a necessary step for obtaining the state in which $\mu^c_L=0$ and for determining solid-liquid phase coexistence. 

\section{Application to EOF-PFC Model}
\label{sec:EOF}

In this section, we apply the thermodynamic relationships for the PFC model developed in the previous section to the EOF-PFC model (see Section \ref{sec:PFC_free_energy}) to demonstrate the procedure for determining phase coexistence, as well as to develop physical intuition about the PFC input parameters. 


\subsection{Free-Energy Density Curves and Phase Coexistence for $bcc$ Fe}
\label{subsec:FED_bcc_Fe}

\begin{figure}[h]
\begin{center}
$
\begin{array}{c}
\includegraphics[height=7cm]{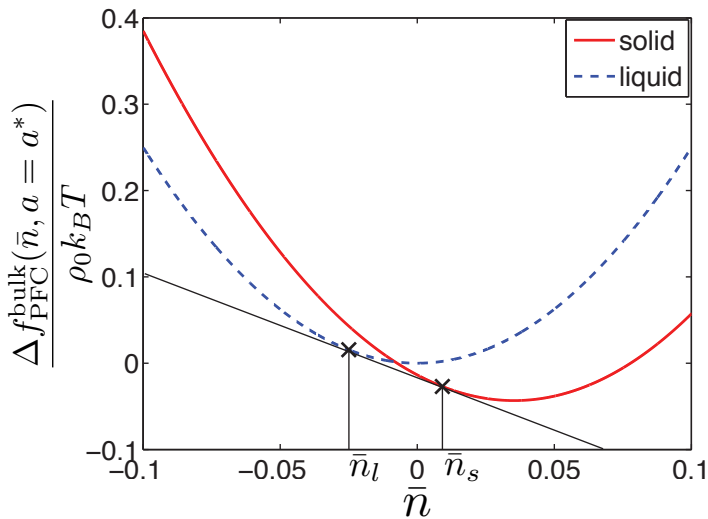} 
\end{array}
$
\end{center}
\caption{\label{fig:f_vs_nbar}  Plot of the dimensionless PFC FEDs of the EOF-PFC model for the solid (solid red line) and liquid (dashed blue line) phases as a function of $\bar{n}$. Each point on the solid FED curve is minimized with respect to lattice spacing, and thus satisfies $\mu^c_L = 0$. The dimensionless coexistence number density for the solid, $\bar{n}_s$, and liquid, $\bar{n}_l$, phases are marked with ``$\times$" marks.}
\end{figure}

We calculate the solid and liquid FED curves for the EOF-PFC model with the fitting parameters of Ref.\ \onlinecite{Jaatinen:2009p2381}, which are listed in Section \ref{sec:PFC_free_energy}. The dimensionless PFC FED for the solid and liquid phases are plotted as functions of $\bar{n}$ in Fig.\ \ref{fig:f_vs_nbar}. 
%
The equilibrium FED for each value of $\bar{n}$ was calculated by relaxing a one-mode approximation of $bcc$ Fe according to the PFC FED via Eq.\ \eqref{dynamics}, as done in Ref.\ \onlinecite{Jaatinen:2010gp}. Each point in the solid FED curve of Fig.\ \ref{fig:f_vs_nbar} satisfies $\mu^c_L=0$, which is achieved by minimizing the PFC FED with respect to lattice spacing. As described in Section \ref{sec:PFC_free_energy}, the value of $a$ that minimizes the PFC FED, denoted as $a^*$, remains constant because the position of the primary peak in the two-body DCF of the PFC model is assumed to be independent of $\bar{n}$. For the calculations in Fig.\ \ref{fig:f_vs_nbar}, $a^*=2.978 \AA$ for all values of $\bar{n}$.  

The scaled dimensionless coexistence number densities for the solid, $\bar{n}_s$, and liquid, $\bar{n}_l$, phases were determined with a common-tangent construction on the FED curves in Fig.\ \ref{fig:f_vs_nbar}. The values for solid and liquid coexistence densities are $\bar{n}_s = 9.17 \times 10^{-3}$ and $\bar{n}_l = -2.49 \times 10^{-2}$, respectively. These values are in agreement with those presented in Ref.\ \onlinecite{Jaatinen:2009p2381}.

\subsection{Diffusion Potential}
\label{subsec:diffusion_potential}

\begin{figure}[h]
\begin{center}
$
\begin{array}{c}
\includegraphics[height=7cm]{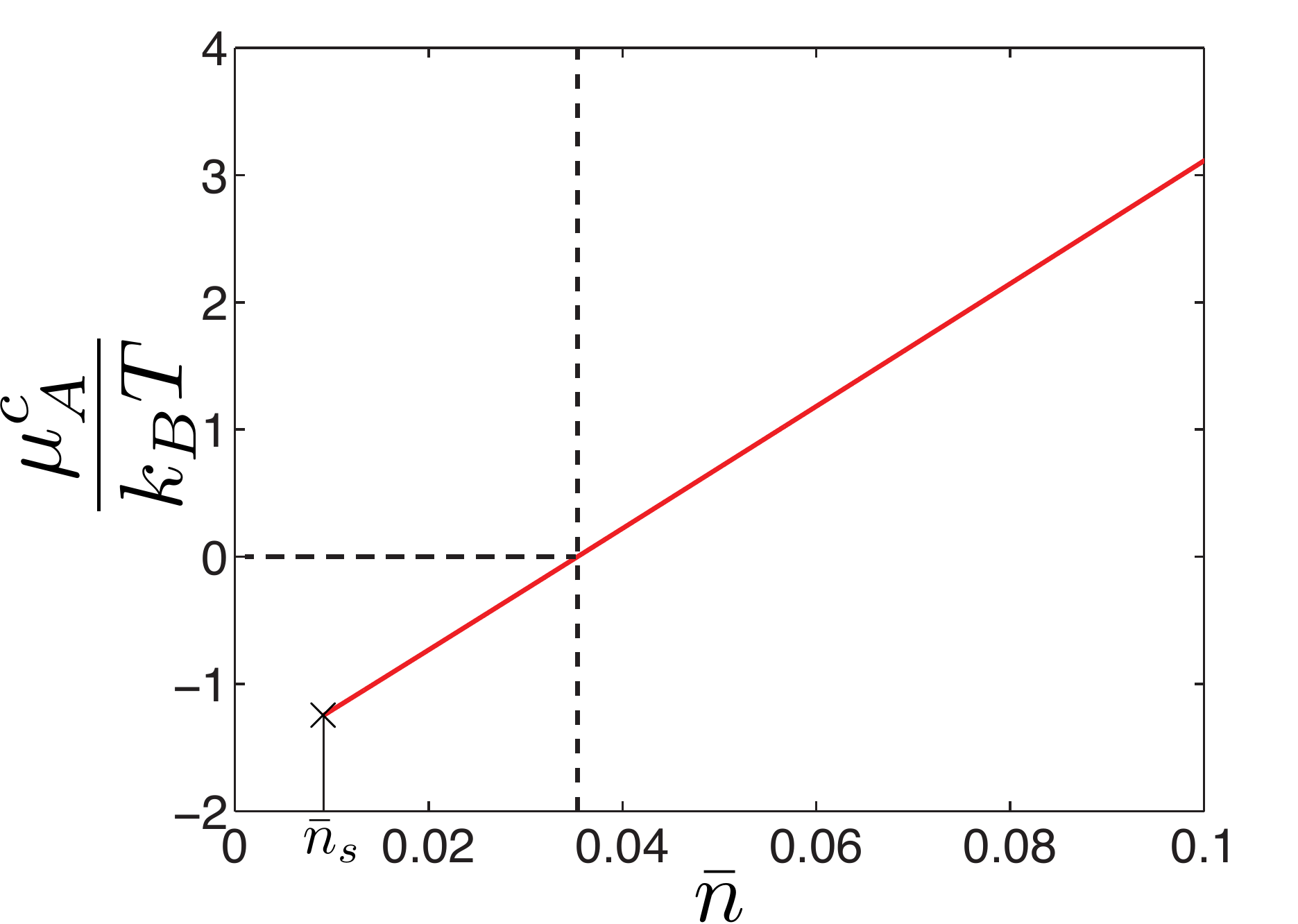} 
\end{array}
$
\end{center}
\caption{\label{fig:MAV}  Plot of diffusion potential, $\mu^c_A$, for $\bar{n}>\bar{n}_s$, where only solid is stable (see Fig.\ \ref{fig:f_vs_nbar}). The dashed vertical line corresponds to $\bar{n}=3.54\times 10^{-2}$, which is the value of $\bar{n}$ where $\mu^c_A=0$ (denoted by horizontal dashed line).}
\end{figure}

Figure \ref{fig:MAV} shows the diffusion potential calculated from Eq.\ \eqref{mu_A^c} for the single-phase solid region of the EOF-PFC model, $\bar{n}>\bar{n}_s$.
%
As can be observed, $\mu^c_A$ transitions from a negative to a positive value at $\bar{n}=3.54\times 10^{-2}$. As described earlier, $\mu^c_A$ represents the energy change due to the addition or removal of atoms in an isothermal system when the number of lattice sites and the crystal volume are held constant. When $\mu^c_A<0$ (left of dashed vertical line in Fig.\ \ref{fig:MAV}), there is a driving force for adding an atom into a vacant lattice site, which decreases as $\bar{n}$ increases. On the other hand, when $\mu^c_A>0$ (right of dashed vertical line in Fig.\ \ref{fig:MAV}), there is a driving force for removing an atom from an occupied lattice site, which decreases as $\bar{n}$ decreases. When $\mu^c_A=0$, there is no driving force for adding or removing atoms to and from lattice sites, and the condition $\mu^c_L=0$ allows us to determine the single-phase equilibrium vacancy density.

In the solid-liquid coexistence region $(\bar{n}_l < \bar{n} < \bar{n}_s)$, chemical equilibrium requires that the chemical potentials of the two phases are equal, which will give rise to  nonzero diffusion potentials. Thus, the equilibrium vacancy density is determined from $\mu^c_A = \mu^l_A$ and $\mu^c_L=0$.

\subsection{An Upper Bound for $\bar{n}$}

As discussed earlier, the number of atoms, vacancies, and lattice sites of a crystal are related to each other by Eq.\ \eqref{N_L}. As a result, the vacancy density, $\rho^c_{V}$, can be expressed in terms of $\rho^c_A$ and $\rho^c_L$ as $\rho^c_V = \rho^c_L - \rho^c_A$. By substituting Eqs.\ \eqref{rhoL} and \eqref{rho^c_A_to_n_bar}, $\rho^c_V$ is expressed in terms of $\bar{n}$ and $\mathcal{V}^c_C$ as
\begin{equation}
\rho^c_V = \frac{\chi^c}{\mathcal{V}^c_C} - \rho_0 \left(\bar{n} + 1\right) .
\label{rhoV_bound}
\end{equation}
%
Since $\mathcal{V}^c_C$ is constant for all $\bar{n}$ values (see Section \ref{subsec:FED_bcc_Fe}), an upper bound for $\bar{n}$ arises when the value $\rho^c_V$ is specified. An upper bound for $\bar{n}$, $\bar{n}_{\text{max}}$, is obtained when $\rho_V^c=0$ (i.e., crystal with no vacancies), 
\begin{equation}
\bar{n}_{\text{max}} = \frac{\chi^c}{\mathcal{V}^c_C \rho_0} - 1 .
\label{nbar_max}
\end{equation}
For $\bar{n}>\bar{n}_{\text{max}}$, the vacancy density takes a negative value, which is unphysical.

The upper bound for the EOF-PFC model with the fitting parameters described in Section \ref{sec:PFC_free_energy} is $\bar{n}_{\text{max}} = -5.46 \times 10^{-2}$, where $\chi^c = 2$ for a $bcc$ structure, $\mathcal{V}^c_C=(2.978 \AA)^3$, and $\rho_0 = 0.0801 \AA^{-3}$. \cite{Jaatinen:2009p2381} Surprisingly, $\bar{n}_{\text{max}} < \bar{n}_l$, where the solid phase is unstable. Therefore, the EOF-PFC model parameterized in Ref.\ \onlinecite{Jaatinen:2009p2381} does not stabilize a $bcc$ Fe structure with $\rho^c_V\ge0$. 



A potential interpretation of $\rho^c_V<0$ is the presence of mobile interstitials. However, further investigation is needed to examine this possibility and its validity. In this work, we simply consider this case as an artifact of the model parameterization and proceed to suggest potential solutions. For example, changing the correlation function, as well as the parameterization of $\Delta f_{id}(n(\textbf{r}))$ in Eq.\ \eqref{free_energy_ideal}, can change the stability of the solid phase such that $\bar{n}_l < \bar{n}_{\text{max}}$. Another approach is to require the position of the primary peak of the two-body DCF, $k_m$, to be a function of $\bar{n}$,
\begin{eqnarray}
k_m(\bar{n}) &=& 2\pi \sqrt{l^2+m^2+n^2} (\mathcal{V}^c_C)^{-\frac{1}{3}} \nonumber\\
&=& 2\pi \sqrt{l^2+m^2+n^2} \left(\frac{\chi^c}{ [\rho^c_V + \rho_0(\bar{n} + 1)]} \right)^{-\frac{1}{3}},
\label{k_m_function}
\end{eqnarray}
where 
\begin{equation}
\mathcal{V}^c_C(\bar{n}) = \frac{\chi^c}{[\rho^c_V + \rho_0(\bar{n} + 1)]} 
\label{V_c_function}
\end{equation}
is obtained by rearranging Eq.\ \eqref{rhoV_bound}, and $l$, $m$, and $n$ are the Miller indices of the primary family of planes (e.g., $l=1$, $m=1$, and $n=0$ for the $bcc$ structure). In this case, the upper bound in $\bar{n}$ no longer arises. Equation \eqref{k_m_function} also allows the direct control of $\rho^c_V$, which has not been previously possible. It also indicates that the manner in which the two-body DCF changes with $\bar{n}$ depends on the crystal structure via $\chi^c$ and the Miller indices. 
Equation \eqref{k_m_function} should only be applied when the change in $\bar{n}$ is due to the addition or removal of an atom while $\theta^c$, $\mathcal{V}^c$, and $N^c_L$ are held constant. This corresponds to changing $\bar{n}$ while holding $\theta^c$ and $\mathcal{V}^c_C$ constant in the PFC model. Note that, in the case where the change in $\bar{n}$ is due to a change in $\mathcal{V}^c$ while $N^c_A$ is held constant, the value of $k_m$ must remain fixed in order to apply a mechanical force, which gives rise to a pressure change.


We point out that a change in $k_m$ reflects a change in the liquid reference state. Therefore, a parameterization that requires $k_m$ to be a function of $\bar{n}$ (as in Eq.\ \eqref{k_m_function}) requires $\rho_0$ and the liquid reference pressure, $P^0$, to change with $\bar{n}$. Since $\bar{n}$ is a function of $\rho_0$, a relationship for $k_m$ as a function $\bar{n}$ can only be obtained from the dependence of $k_m$ on $\rho_0$, which must be determined from atomistic simulations.


\section{Summary and Discussion}
\label{sec:conclusion}

In this paper we have applied the thermodynamic formalism for crystalline solids of Larch\'e and Cahn \cite{Larche:1973wp} to develop a thermodynamic relationship between the PFC free energy and thermodynamic state variables. This relationship allows us to examine the thermodynamic processes associated with varying the PFC model parameters. We showed that varying $\bar{n}$ while keeping the unit-cell volume, $\mathcal{V}^c_C$ (and thus lattice spacing, $a$), and temperature, $\theta^c$, of the bulk crystalline phase constant in a PFC simulation reflects the thermodynamic process of adding or removing atoms to and from lattice sites. Furthermore, changing the computational size of a PFC simulation, while keeping $\bar{n}$ and $\theta^c$ constant, reflects the thermodynamic process of adding or removing lattice sites.

The equilibrium conditions between bulk crystalline solid and liquid phases were then imposed on the thermodynamic relationships for the PFC model to obtain a procedure for determining solid-liquid phase coexistence, which we found to be in agreement with the method commonly used in the PFC community. By using the procedure, we found that no stable $bcc$ phase with a vacancy density greater than or equal to zero exists for the EOF-PFC model that has been parameterized to $bcc$ Fe \cite{Jaatinen:2009p2381}. Therefore, we proposed an alternative parameterization of the EOF-PFC model, which requires the position of the primary peak of the two-body DCF to be a function of $\bar{n}$. The implementation of this parameterization will be a topic of future work.
 
Although several thermodynamic processes associated with changing PFC input parameters (i.e., $\bar{n}$ and $a$) were elucidated for bulk phases in this work, a thermodynamic framework for interfaces in the PFC model remains to be developed. An extension of this work that considers interfaces is needed in order to gain a rigorous, quantitative understanding of PFC simulation results that contain interfaces and grain boundaries. Furthermore, since we do not consider the equilibrium conditions between different crystal phases, the use of the common-tangent construction to determine solid-solid phase coexistence \cite{Elder:2004vn,Jaatinen:2010gp,Greenwood:2011bs} remains to be verified.  The extension of this work to systems containing interfaces and different crystal phases will be the topics of future investigations.

\section{Acknowledgements}

This research was supported by National Science Foundation (NSF) under grant No.\ DMR-1105409. The calculations in this work was made possible by the services provided by Advanced Research Computing at the University of Michigan, Ann Arbor. V. W. L. Chan is indebted to David Montiel, Jason Luce, and Peter Voorhees for numerous discussions and helpful comments on this topic. The authors would also like to thank Elizabeth Hildinger for her help with editing the manuscript. 

\appendix
\section{Gibbs-Duhem Relation for Bulk Crystalline Solid}
\label{Appendix:Gibbs_Duhem}
We derive the Gibbs-Duhem relation for a bulk crystalline solid following Voorhees and Johnson \cite{Voorhees:2007tg}. The internal energy of a one-component crystal is a function of $S^c$, $\mathcal{V}^c$, $N^c_A$, and $N^c_L$,
\begin{equation}
E^c = E^c(S^c,\mathcal{V}^c,N^c_A,N^c_L).
\label{E^c}
\end{equation}
The differential form of $E^c$ is
\begin{equation}
dE^c = \theta^c d S - P^c d\mathcal{V}^c + \mu_A^c dN^c_A + \mu_L^c dN^c_L,
\label{dE^c}
\end{equation}
where 
\begin{equation}
\theta^c \equiv -\frac{\partial E^c}{\partial S^c} \bigg |_{\mathcal{V}^c, N_A^c, N_L^c}, \ \ \ P^c \equiv -\frac{\partial E^c}{\partial \mathcal{V}^c} \bigg |_{S^c, N_A^c, N_L^c}, \ \ \ \mu_A^c \equiv \frac{\partial E^c}{\partial N_A^c} \bigg |_{S^c, \mathcal{V}^c, N_L^c}, \ \ \ \mu_L^c \equiv \frac{\partial E^c}{\partial N_L^c} \bigg |_{S^c, \mathcal{V}^c, N_A^c} .
\end{equation}
Since Eq.\ \eqref{E^c} is a homogenous function of degree one, Eq.\ \eqref{dE^c} yields
\begin{equation}
E^c(S^c,\mathcal{V}^c,N_A^c,N_L^c) =  S^c \theta^c -P^c\mathcal{V}^c + \mu_A^c N_A^c + \mu_L^c N_L^c.  
\label{E^c_euler}
\end{equation}
Differentiating Eq.\ \eqref{E^c_euler} and subtracting from Eq.\ \eqref{dE^c} gives us the Gibbs-Duhem relation for a bulk crystalline solid,
\begin{equation}
0 = S^c d\theta^c -\mathcal{V}^cdP^c +N_A^c d\mu_A^c + N_L^c d\mu_L^c.
\end{equation}

%

\newpage
\bibliographystyle{apsrev}
\bibliography{PFC_thermo}

\end{document}